\documentclass[10pt,twocolumn]{article} 
\usepackage{simpleConference}
\usepackage{times}
\usepackage{graphicx}
\usepackage{amssymb}
\usepackage{url,hyperref}

\usepackage{graphicx}   % For including graphics
\usepackage{subcaption} % For subfigures

\usepackage{fancyhdr}
\usepackage{lipsum} % For sample text

\fancypagestyle{firstpagefooter}{
    \fancyhf{}

    \fancyfoot[C]{© {Owner/Author | ACM} {2024}. This is the author's version of the work. It is posted here for your personal use. Not for redistribution. The definitive Version of Record was accepted by {Sigcomm 2024 Poster}.
    }
}

\begin{document}

%-------------------------------------------------- Title -----------------------------------------------------%

\title{Poster: Flexible Scheduling of Network and Computing Resources for Distributed AI Tasks}%

%------------------------------------------------- Authors-----------------------------------------------------%

\author{Ruikun Wang$^{\ast}$, Jiawei Zhang$^{\ast}$, Qiaolun Zhang$^{\dagger}$, Bojun Zhang$^{\ast}$, Zhiqun Gu$^{\ast}$, Aryanaz Attarpour$^{\dagger}$, \\ Yuefeng Ji$^{\ast}$, Massimo Tornatore$^{\dagger}$\\
% Jiaheng Xiong$^{(1)}$,  Qiaolun Zhang$^{(1)^{*}}$, Ruikun Wang$^{(2)}$, Alberto Gatto$^{(1)}$, \\Francesco Musumeci$^{(1)}$, Massimo Tornatore$^{(1)}$\\
$^{\ast}$Beijing University of Posts and Telecommunications, Beijing, China  \\
$^{\dagger}$Politecnico di Milano \\
Corresponding author: Zhiqun Gu and Jiawei Zhang
 ({guzhiqun,zjw}@bupt.edu.cn)
}

%\author{
%    Jiaheng Xiong\textsuperscript{(1)}, 
%    Qiaolun Zhang\textsuperscript{(1)}\textsuperscript{*}, 
%    Ruikun Wang\textsuperscript{(2)},
%    Alberto Gatto\textsuperscript{(1)}, 
%    Francesco Musumeci\textsuperscript{(1)}, \\
%    Massimo Tornatore\textsuperscript{(1)}\\
%    \textsuperscript{(1)} Politecnico di Milano, Milan, Italy, Corresponding author: *{\uline{qiaolun.zhang@mail.polimi.it}}
%   , \\
%   \textsuperscript{(2)} Beijing University of Posts and Telecommunications, Beijing, China 
%}

\maketitle

% Apply the footer to the first page
\thispagestyle{firstpagefooter}

\begin{abstract}
  Many emerging Artificial Intelligence (AI) applications require on-demand provisioning of large-scale computing, which can only be enabled by leveraging distributed computing services interconnected through networking. To address such increasing demand for networking to serve AI tasks, we investigate new scheduling strategies to improve communication efficiency and test them on a programmable testbed. We also show relevant challenges and research directions.

\end{abstract}

% \keywords{Artificial Intelligence, Networking}

  %  50\% reduction in spectrum occupation compared to single-mode transmission and Full MIMO approach, and x\% reduction compared to dedicated path protection.

%-------------------------------------------------- Introduction Section -------------------------------------------------------%

\section{Introduction}
    
Artificial Intelligence (AI) has become a key technology for many scientific communities~\cite{10.1145/3603269.3610855}, with a broad set of applications, such as %network congestion control~\cite{10.1145/3603269.3604838}, 
computer vision and natural language processing~\cite{He2015DeepRL,10.5555/3295222.3295349}. 
In particular, distributed AI is an attractive paradigm, as it can provide an efficient collaboration of different local AI models, typically leveraging privacy-preserving communication to aggregate information~\cite{10239353,10239286,10234534}. 
These distributed AI tasks represent a new data-transfer service type to be supported in telecom/cloud networks. The local AI models need to be interconnected via large bandwidth pipes to support their needs for continuous synchronization and update of the models~\cite{10.1145/3603269.3604835,10.1145/3603269.3604869}. In particular, with the emergence of generative AI~\cite{10.1145/3603269.3610856}, the model size is increasing rapidly, leading to much greater communication overhead and latency. 
Thus, novel methodologies for cooperative/flexible scheduling of network and computing resources need to be investigated to support these distributed AI tasks.
While some initial works have already appeared~\cite{9606720,10038471,13235454}, several fundamental questions remain open: 
(1) \textbf{how can we model the computing and networking requirements of distributed AI tasks?} Such requirements can be expressed, e.g., in terms of \textit{model training, aggregating and communication latency}, and \textit{bandwidth demand}. Note that AI tasks can be implemented using different machine learning (ML) models that include different parameters %and training data
~\cite{9083410}.
(2) \textbf{How to strategically (re)schedule network and computing resources?} \textit{Routing paths} and \textit{aggregation procedures} must be initially scheduled for each AI task, and then re-scheduled when the deployed AI tasks and networks change. Predictability of training iteration can be leveraged to optimize scheduling.

\begin{figure}[tp]
\centering
\includegraphics[width=0.88\linewidth]{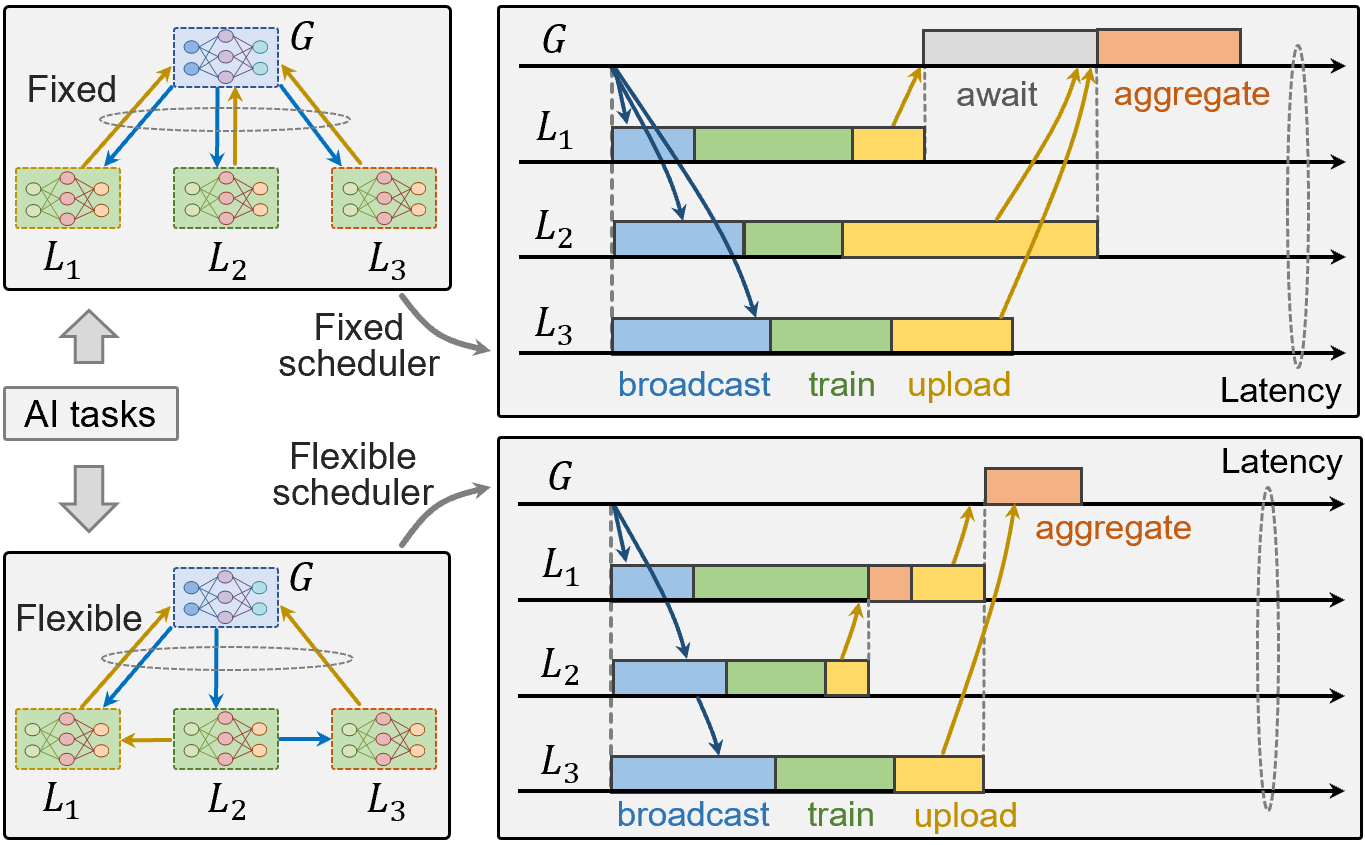}
\vspace{-0.3cm}
\caption{A comparison of fixed and flexible scheduler.}
\label{label-figure1}
\vspace{-0.54cm} 
\end{figure}

In this poster, we aim to take a step towards flexible scheduling of network and computing resources for distributed AI tasks.
We propose a strategy based on the minimum spanning tree (MST) that dynamically decides routing paths and aggregation operations. 
We test it using a programmable orchestrator that manages heterogeneous resources in logically centralized manner.
Due to limitation on testbed, we take a fixed scheduler using shortest path and first fit (SPFF) as baselines~\cite{13235454}, while the comparison with stronger baselines will come as future works. We implement them for several AI-task use cases on a real testbed, showing its superiority in latency and bandwidth saving compared to baselines.

\section{Experimental Framework}
As shown in Figure.~\ref{label-figure1}, the fixed scheduler considers a fixed set of direct communication links between the global model ($G$) and each local model (e.g., $L_1, L_2, L_3$). AI model weights are transmitted using end-to-end links in broadcast ($G \rightarrow L_i$) and upload ($L_i \rightarrow G$) procedures\footnote{AI weights are transmitted from the global model to local models in broadcast procedure, while upload is an opposite communication procedure.}, and then only aggregated in the node with a global model %, which causes a waste of bandwidth and high latency
~\cite{13235454}.
In contrast, the % our proposed 
flexible scheduler finds a suitable connectivity set, e.g., $G \rightarrow L_1$, $G \rightarrow L_2 \rightarrow L_3$ for the broadcast procedure, and further schedules routing paths and aggregation operations. 
We first build auxiliary graphs for broadcast and upload procedures, 
respectively. 
We initialize each link of the broadcast/upload graphs according to bandwidth consumption and latency (if AI tasks pass through the link), and then find MSTs between the global model and local models. The links of MSTs are considered as routing paths, and the aggregation operations happen in the middle and final nodes of upload procedure.

\begin{figure}[tp]
\centering
\includegraphics[width=0.94\linewidth]{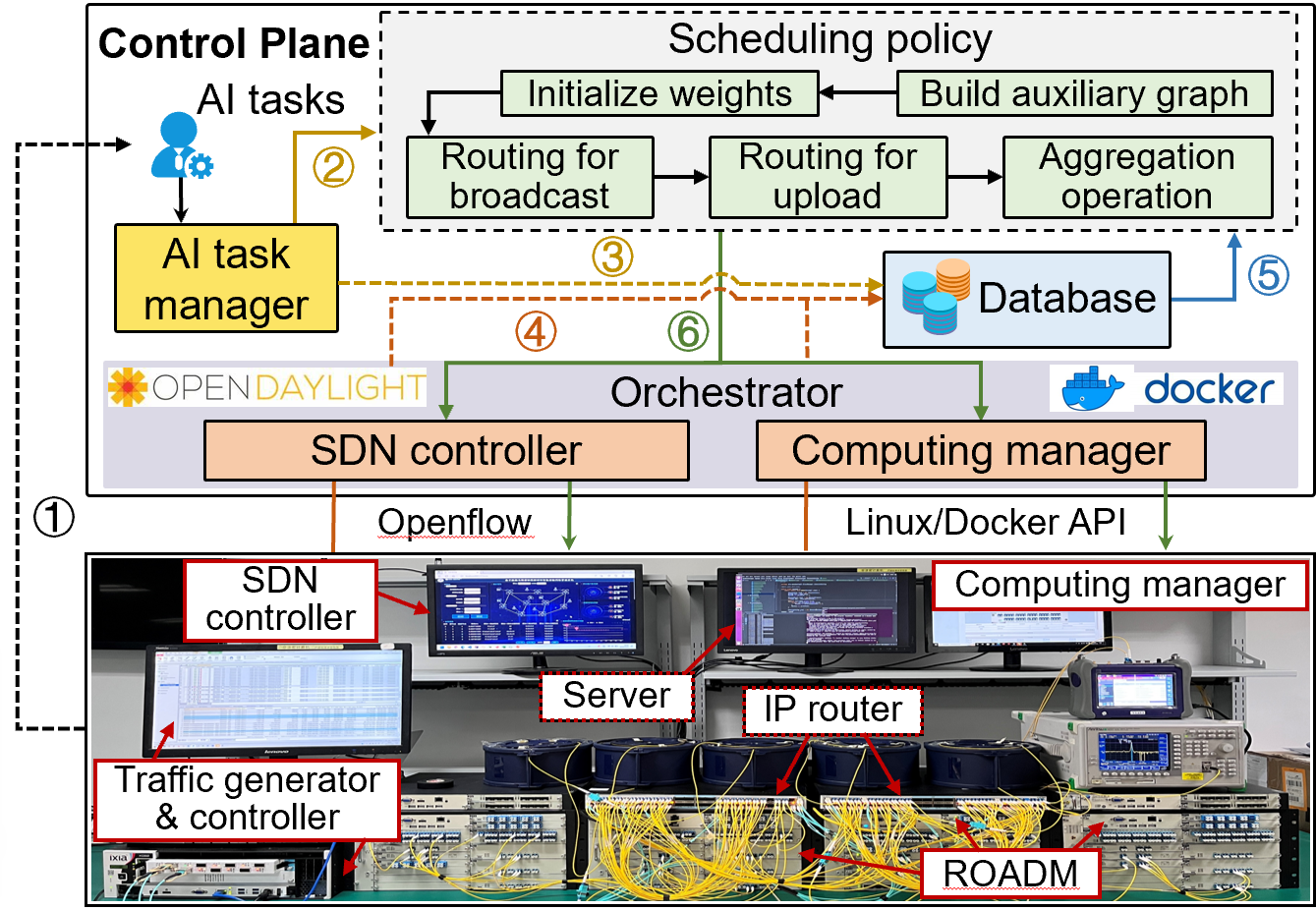}
\vspace{-0.25cm}
\caption{Experimental framework and procedures.}
\label{label-figure2}
\vspace{-0.6cm}
\end{figure}

The experimental framework is depicted in Figure.~\ref{label-figure2}~\cite{10477618,10025402}. 
Reconfigurable optical add/drop multiplexers (ROADM) and IP routers are used for traffic switching and grooming, and live traffic is injected by a traffic generator.
Linux OS and dockers are deployed in several servers to support AI tasks.
They are controlled by an SDN controller and a computing manager. 
An orchestrator is used to report networking conditions to the database, and configure routing paths according to the scheduling policy. An AI task manager is responsible for managing new AI tasks and storing them into database. The scheduling policy is also embedded into the testbed.

\section{Evaluations}

\begin{figure}[tbp]
    \centering
    \begin{subfigure}[b]{0.44\textwidth}
        \centering
        \includegraphics[width=0.95\textwidth]{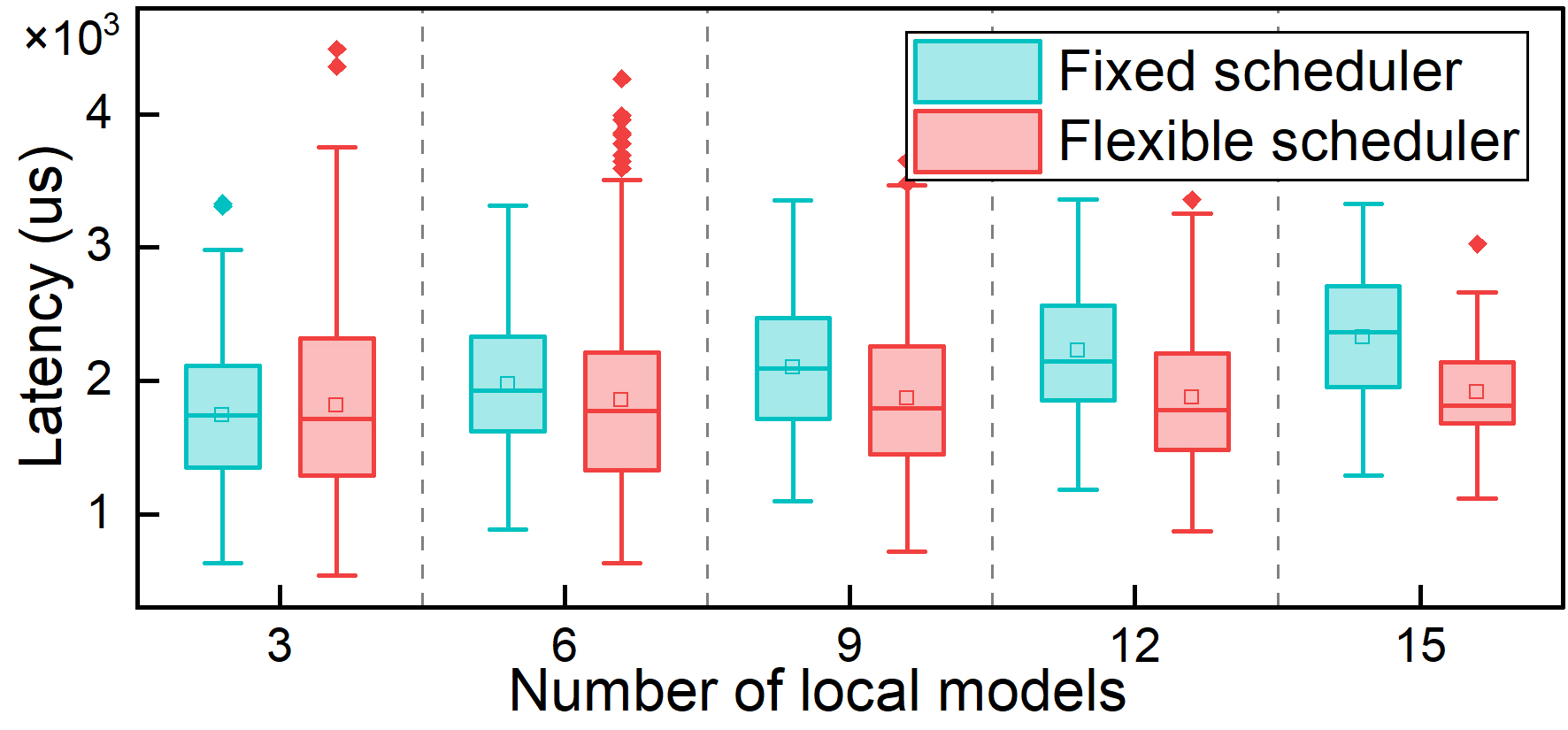}
        \vspace{-0.18cm}
        \caption{Latency vs. number of local models.}
        \label{label-figure3-1}
    \end{subfigure}
    % \hfill
    \begin{subfigure}[b]{0.44\textwidth}
        \centering
        \includegraphics[width=0.95\textwidth]{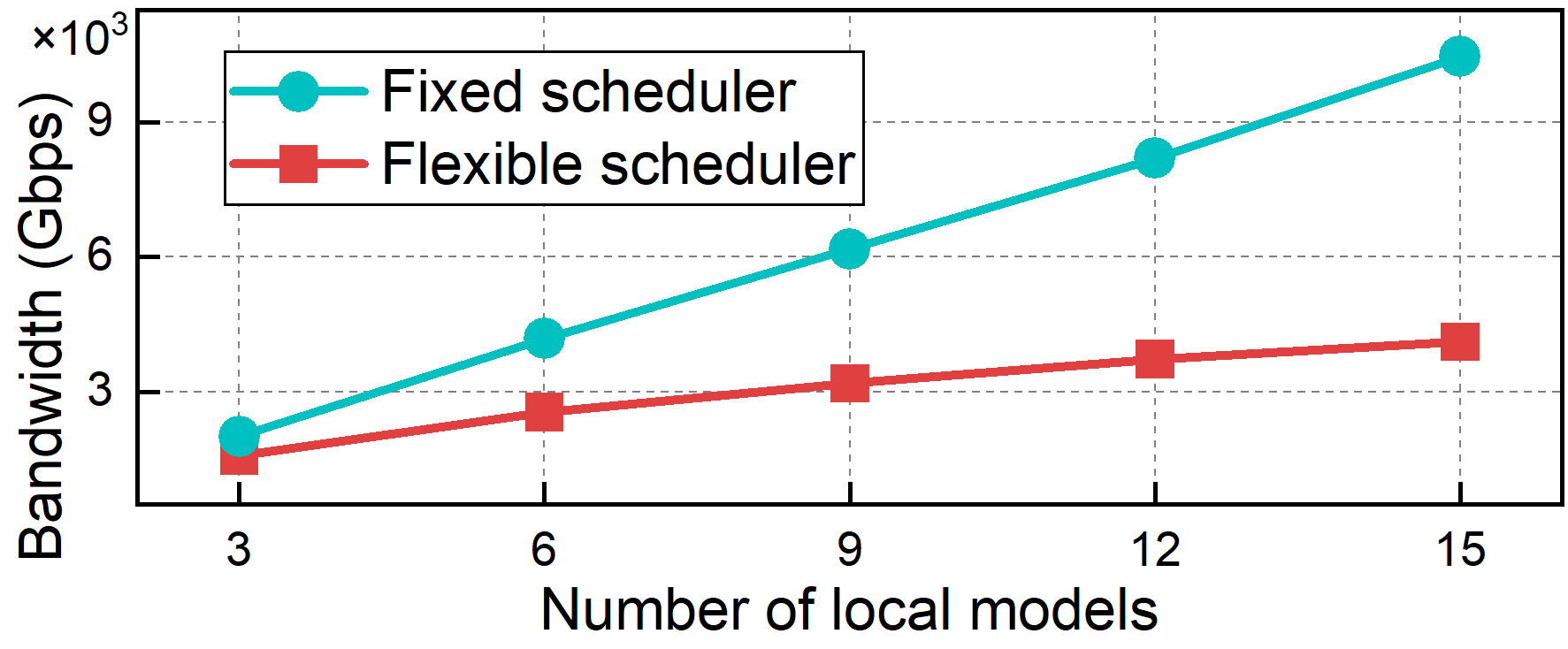}
        \vspace{-0.18cm}
        \caption{Bandwidth vs. number of local models.}
        \label{label-figure3-2}
    \end{subfigure}
    \vspace{-0.35cm}
    \caption{Evaluation results.}
    \label{label-figure3}
    \vspace{-0.6cm}
\end{figure}

We generate 30 AI tasks to evaluate the proposed scheduling policy~\cite{13235454}.
Figure.~\ref{label-figure3-1} shows the total latency (both model training and communication) for an increasing number of local models, showing that the flexible scheduler finishes model training incurring much lower latency. 
For example, when the number of local models is 15, the average latency is $1.9$ $ms$ and $2.3$ $ms$ for different schedulers.
Figure.~\ref{label-figure3-2} reports the consumed bandwidth. The fixed scheduler shows a nearly linear relationship between consumed bandwidth and the number of local models, as it needs to build a large amount of end-to-end routing paths. However, the proposed scheduler consumes less bandwidth, since AI tasks can use some existing paths to transmit model weights. The above results show that useful scheduling strategies can improve communication efficiency for AI tasks.

\section{Open Challenges}

\noindent \textbf{\#1: novel scheduling strategies.} In distributed AI systems, each local model contributes to the global model based on its local data. Thus, we should strategically select only those local models containing useful data to improve model learning%, hence reducing bandwidth and latency
. We also need to balance a trade-off between re-scheduling (temporary interruption% of AI tasks
) and bandwidth/latency saving.

\noindent \textbf{\#2: novel communication protocols.} %When transmitting AI tasks, 
TCP/IP protocols consume a lot of CPU resources and packet heads, which reduces communication/training efficiency. A \textit{protocol based on remote direct memory access (RDMA)} is needed for direct communication between buffers~\cite{8895760,10.1145/3603269.3604849,ichikawa2021rdma}, while several challenges also remain:
\textit{i)} how to achieve \textit{near zero packet loss} for reducing the amount of re-transmitted packet;
\textit{ii)} how to deal with \textit{performance degradation in long-distance networks}.

\noindent \textbf{\#3: novel network architectures.} 
The existing network architectures are usually designed to connect access/metro/core nodes, which is not suitable to connect distributed computing nodes. An \textit{all-optical network based on spine-leaf architectures} is needed to provide large-bandwidth and low-latency pipelines~\cite{11dfetbtrbr}. However, we need to further study how to \textit{collaboratively manage optical wavelengths and timeslots}.

\section*{Acknowledgements}
Zhiqun Gu and Jiawei Zhang are the corresponding authors (\{guzhiqun,zjw\}@bupt.edu.cn). This work was supported by National Key R\&D Program of China (2022YFB2903700), and the Italian Ministry of University and Research (MUR) and the European Union (EU) under the PON/REACT project.

\bibliographystyle{abbrv}
\bibliography{refs}

%%%%%%%%%%%%%%%%%%%%%%%%%%%%%%%%%%%%%%%%%%%%%
%---------------------------------------------- End of Document -----------------------------------------------%
\end{document}